\documentclass[prl,showpacs,twocolumn]{revtex4}
\usepackage{graphicx}
\usepackage[latin1]{inputenc}
\usepackage{amsmath}

\newcommand{\eq}[1]{Eq.~(\ref{#1})}
\newcommand{\fig}[1]{Fig.~\ref{#1}}

\newcommand{\bit}{\begin{itemize}}
\newcommand{\eit}{\end{itemize}}

\newcommand{\etal}{{\em et al.}}

\begin{document}
\title
{Electronic structure of strongly correlated $d$-wave superconductors}
\author{Bernhard Edegger$^{1,2,3}$, V.~N.~Muthukumar$^2$,
Claudius Gros$^1$, and P.~W.~Anderson$^3$}
\affiliation{$^1$ Institute for Theoretical Physics, Universit\"at
Frankfurt, D-60438 Frankfurt, Germany} \affiliation{$^2$
Department of Physics, City College of the City University of New
York, New York, NY 10031} \affiliation{$^3$ Department of Physics,
Princeton University, Princeton, NJ 08544}
\date{\today}
\begin{abstract}
We study the electronic structure of a strongly correlated
$d$-wave superconducting state. Combining a renormalized mean
field theory with direct calculation of matrix elements, we obtain
explicit analytical results for the nodal Fermi velocity, $v_F$,
the Fermi wave vector, $k_F$, and the momentum distribution,
$n_k$, as a function of hole doping in a Gutzwiller projected
$d$-wave superconductor. We calculate the energy dispersion,
$E_k$, and spectral weight of the Gutzwiller-Bogoliubov
quasiparticles, and find that the spectral weight associated with
the quasiparticle excitation at the antinodal point shows a non
monotonic behavior as a function of doping. Results are compared
to angle resolved photoemission spectroscopy (ARPES) of the high
temperature superconductors.
\end{abstract}
\maketitle

Recent progress in angle resolved photoemission spectroscopy
(ARPES) of the high temperature superconductors has led to
considerable interest in the electronic structure of a strongly
correlated $d$-wave superconducting state \cite{campuzano_rev,
shen_rev}. Experiments show a variety of interesting phenomena and
it is generally agreed that strong electronic correlations play a
dominant role in explaining some of the universal spectral
features. Hence it would be desirable to obtain, \textit{e.g.},
for comparison with experimental results, explicit analytical
results from a simple model of a strongly correlated $d$-wave
superconductor.

Thus motivated, we consider in this paper, the electronic
structure of a Gutzwiller projected superconductor, defined by the
ground state wave function, $\exp(iS)|\Psi \rangle \equiv
\exp(iS)P_G |\Psi_0\rangle$. The projection operator, $P_G \equiv
\prod_i (1-n_{i\uparrow} n_{i\downarrow})$, acting on a BCS wave
function, $|\Psi_0\rangle$, eliminates states with double
occupancies in $|\Psi \rangle$. The operator, $\exp(iS)$, allows
for a systematic calculation of corrections to the fully projected
state, $|\Psi\rangle$ \cite{paramekanti,randeria_05}. Gutzwiller
projected states, $|\Psi \rangle$, were initially proposed as
variational states to describe superconductivity in the proximity
of a Mott insulating phase \cite{pwa_87,zhang_88,VMC}. They have
been used recently for numerical investigations of the electronic
structure of the high temperature superconductors. Paramekanti
\textit{et al.}, used Variational Monte Carlo (VMC) to study some
spectral properties of a Gutzwiller projected $d$-wave
superconductor \cite{paramekanti}. Yunoki \textit{et al.},
extended the VMC technique to the direct calculation of excited
states in Jastrow-Gutzwiller wave functions \cite{yunoki_05}. The
VMC technique also allows to study the coexistence of
superconductivity with antiferromagnetism \cite{shih04}, and the
quasiparticle current renormalization \cite{ivanov05}.

In this paper, we will follow an alternate route and use a
combination of renormalized mean field theory (RMFT)
\cite{zhang_88} and direct calculation of matrix elements
\cite{mxelem,randeria_05} to examine the electronic structure of a
Gutzwiller projected $d$-wave superconductor. Though Gutzwiller
projection is only an approximate method to treat the effects of
strong correlations, the advantage lies in its directness and
clarity. Explicit analytical expressions derived in this paper can
be used to evaluate directly the successes as well the limitations
of this approach. Using RMFT, we determine the energy dispersion,
$E_k$, of the Gutzwiller-Bogoliubov quasiparticles, and the nodal
Fermi velocity, $v_F$. We show that $v_F$ stays finite as the
(Mott) insulating phase is approached. We calculate the spectral
weight associated with the $d$-wave quasiparticles, and the
momentum distribution, $n_k$, by a direct evaluation of the
relevant matrix elements \cite{mxelem}. We find that the
quasiparticle weight associated with the antinodal excitation
exhibits a non monotonic behavior as a function of doping.

We consider the one band Hubbard model,
$$
H=\,-\sum_{\langle ij\rangle,\sigma} t_{(ij)} \left(
c_{i\sigma}^\dagger c_{j\sigma} + c_{j\sigma}^\dagger c_{i\sigma}
\right) \,+ \,U\,\sum_i\,n_{i\uparrow} n_{i\downarrow}\ ,
$$
where the hopping integrals, $t_{(ij)}$, connect sites $i$ and
$j$. We will restrict our attention to nearest ($t$), and next
nearest ($t^\prime=-t/4$) neighbor hopping. We choose an onsite
repulsion $U=12\,t$; \textit{i.e.}, we work in the strong coupling
regime $U\gg t,t'$ \cite{dagotto_94}. In this limit, a unitary
transformation, $\exp({iS})$, yields an effective Hamiltonian,
$H_{\rm eff}=\exp({-iS}) H \exp({iS})$, which is block diagonal
and does not mix states between the lower and upper Hubbard bands
\cite{gros_87,paramekanti}. To $\mathcal{O}(t^2/U)$, it corresponds to
the $t-J$ Hamiltonian with the correlated hopping term (three-site
term), for which the Gutzwiller wave function $P_G|\Psi_0\rangle$
has been established as an excellent variational ground state
\cite{VMC}. Retransformation of the trial wave function,
$\exp(iS)P_G |\Psi_0\rangle$, then provides a systematic way to
study the Hubbard model using Gutzwiller projected states.

Two steps are necessary to obtain explicit analytic expressions
for the low energy properties in the strong coupling regime. The
first is the Gutzwiller renormalization procedure, where the
effects of the projection $P_G$ are taken into account by
appropriate renormalization factors, following Hilbert space
counting arguments: $\langle \Psi_0 | P_G H_{{\rm eff}} P_G
|\Psi_0 \rangle \approx \langle \Psi_0 |\tilde{H}_{{\rm
eff}}|\Psi_0 \rangle$ \cite{zhang_88,edegger_05,GutzRef}.

The next step is the realization that $\tilde{H}_{{\rm eff}}$
allows for two types of molecular-fields:
the hopping amplitude, $\xi_{r}\equiv\sum_\sigma\, \langle
c^\dagger_{i\sigma} c_{i+r\sigma} \rangle_0$,
and the singlet pairing amplitude, $\Delta_{r}\equiv\,\langle
c^\dagger_{i\uparrow} c^\dagger_{i+r\downarrow}
-c^\dagger_{i\downarrow} c^\dagger_{i+r\uparrow} \rangle_0$,
where $r$ connects nearest (nn) or
next nearest neighboring (nnn) sites;
$\langle...\rangle_0$ denotes the expectation value with respect
to $|\Psi_0 \rangle$.

This decoupling scheme of the renormalized
Hamiltonian (RMFT) leads to a BCS ground state
$|\Psi_0 \rangle = \prod_k (u_k + v_k
c^\dagger_{k\uparrow}c^\dagger_{-k\downarrow})~|0\rangle$,
with $v_k^2=(1-{\zeta_k}/{E_k})/2$
and $u_k^2=1-v_k^2$. The corresponding
gap equations are
$\xi_r=-{1}/L \sum_k \cos (k\,r) {\zeta_k}/{E_k}$
and
$\Delta_r=1/L \sum_k \cos (k\,r) {\tilde\Delta_k}/{E_k} $,
together with the condition
$x= {1}/L \sum_k {\zeta_k}/{E_k}$
for the hole-doping concentration
\cite{zhang_88,calc_energy}.

Solving the gap equations, we
find that a $d$-wave pairing state is most stable for $x\leq0.4$.
In this case, $\Delta\equiv|\Delta_x|=|\Delta_y|$ with
$\Delta_x=-\Delta_y$, $\xi\equiv\xi_x=\xi_y$, and
$\xi'\equiv\xi_{x+y}=\xi_{x-y}$. The dispersion relation of the
Gutzwiller-Bogoliubov quasiparticle is given by,
$E_k = \sqrt{\zeta_k^2 + \Delta^2_k}$,
where,
\begin{eqnarray}
\zeta_k &=& -\left(2g_t t+ J\,\frac \xi 4 \,x_1\,+J'\,\frac{\xi'} 4\,x_2\right)
(\cos k_x+\cos k_y)\nonumber \\
& &-\left(2g_t t'+ J''\,\frac {\xi'} 4 \,x_1\,+J'\,\frac{\xi}
4\,x_2\right)\,2\,\cos
k_x \cos k_y \nonumber\\
& &-x_D \sum_{\tau\neq\tau'} \frac{t_\tau t_{\tau'}}{4U}
\cos\left[k (\tau-\tau') \right]\,-\mu \ , \label{epsk}\\
\tilde \Delta_ k &=& J\,\frac \Delta 4\,[3g_s+1-(3+x)g_3]\,(\cos k_x-\cos
k_y)~. \label{delk}
\end{eqnarray}
In \eq{epsk} and \eq{delk}, the Gutzwiller factors corresponding
to the kinetic, superexchange and the three-site term are
$g_t=2x/(1+x)$, $g_s=4/(1+x)^2$ and $g_3=4x/(1+x)^2$ respectively
\cite{edegger_05}. The last sum in \eq{epsk} is a sum over all
pairs of  neighboring sites $\tau$ and $\tau^\prime$, where
$t_\tau$ and $t_{\tau'}$ are the nn and nnn hopping terms. We
define, $J=4t^2/U$, $J'=4t't/U$, and $J''=4t'^2/U$ and abbreviate,
$x_1=3g_s-1+3(3-x)g_3$, $x_2=4(3-x)g_3$, and $x_D=(1-x^2)g_3$ in
\eq{epsk}.

\begin{figure}[b]
  \centering
 \includegraphics*[width=0.48\textwidth]{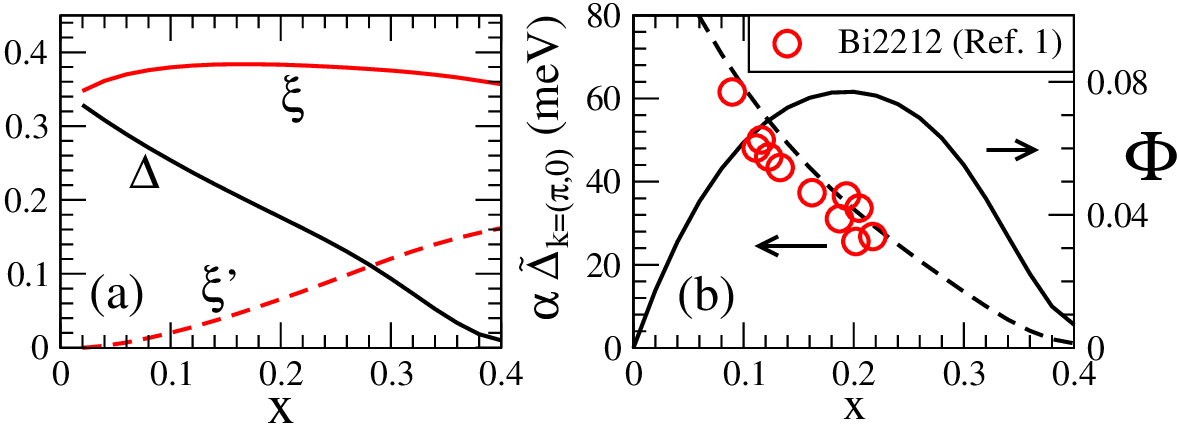}
 \caption{(color online) (a) Doping dependence of the dimensionless
mean field parameters $\xi$, $\xi'$, $\Delta$; (b) Doping
dependence of (solid) the SC order parameter, $\Phi$, and (dashed)
the SC gap, $\alpha\,|\tilde \Delta_k|$, at $k=(\pi,0)$ for $t=300$ meV.
The RMFT SC gap is scaled by a factor $\alpha=1/2$ for
comparison with experimental data (red circles, Bi2212 \cite{campuzano_rev})
}
\label{phasediagram}
\end{figure}
\fig{phasediagram}(a) shows the doping dependence of the mean
field hopping parameters, $\xi$ and $\xi'$, and the $d$-wave mean
field pairing, $\Delta$, obtained by solving the RMFT gap
equations. Since $g_t,\ g_3,\ x_2,\ x_D$ and $\xi'$ vanish
as $x\rightarrow 0$,
the dispersion relation,
$\zeta_k\to-(11/4)J\xi(\cos k_x+\cos k_y)$ in the
limit of zero doping.

The $d$-wave order parameter $\Delta$ decreases nearly
linearly with doping and vanishes around $x\approx0.4$, for our
choice of parameters. This is in agreement with previous
studies \cite{zhang_88,kotliar_88}. Note that we retain
\emph{all} terms of $\mathcal{O}(t^2/U)$ in the effective Hamiltonian,
$H_{\rm eff}$, including the three-site
term, which suppresses
$d$-wave pairing in the overdoped regime.

The doping dependence of the superconducting (SC) gap, $|\tilde
\Delta_k|$ at $k=(\pi,0)$, is shown in \fig{phasediagram}(b). The
doping dependence
resembles experimental observations quite well. However, the gap is
overestimated by a factor of about 2 [see scaling factor $\alpha = 1/2$ in
\fig{phasediagram}(b)]
within mean field theory which neglects additional
off site correlations as well as dynamical effects due to the
motion of holes \cite{pwa_02}. The SC gap is not identical to the
true SC order parameter, $\Phi\equiv|\langle c^\dagger_{i\uparrow}
c^\dagger_{i+\tau\downarrow}-c^\dagger_{i\downarrow}
c^\dagger_{i+\tau\uparrow}\rangle|$ \cite{zhang_88,paramekanti}.
Here, $\tau$ is a neighboring site and $\langle...\rangle$
represents the expectation value calculated with the retransformed
wave function, $\exp(iS)P_G |\Psi_0\rangle$. Calculating $\Phi$ to
$\mathcal{O}(t/U)$, we find, $\Phi \approx \,g_t \Delta +
(t/U)\,g_3\,{(6-x)}\,\Delta\,\xi$, where we set $t' \approx 0$,
for simplicity. As shown in  \fig{phasediagram}(b),
$\Phi$ vanishes as $x \rightarrow 0$, as it should.
\begin{figure}
  \centering
 \includegraphics*[width=0.48\textwidth]{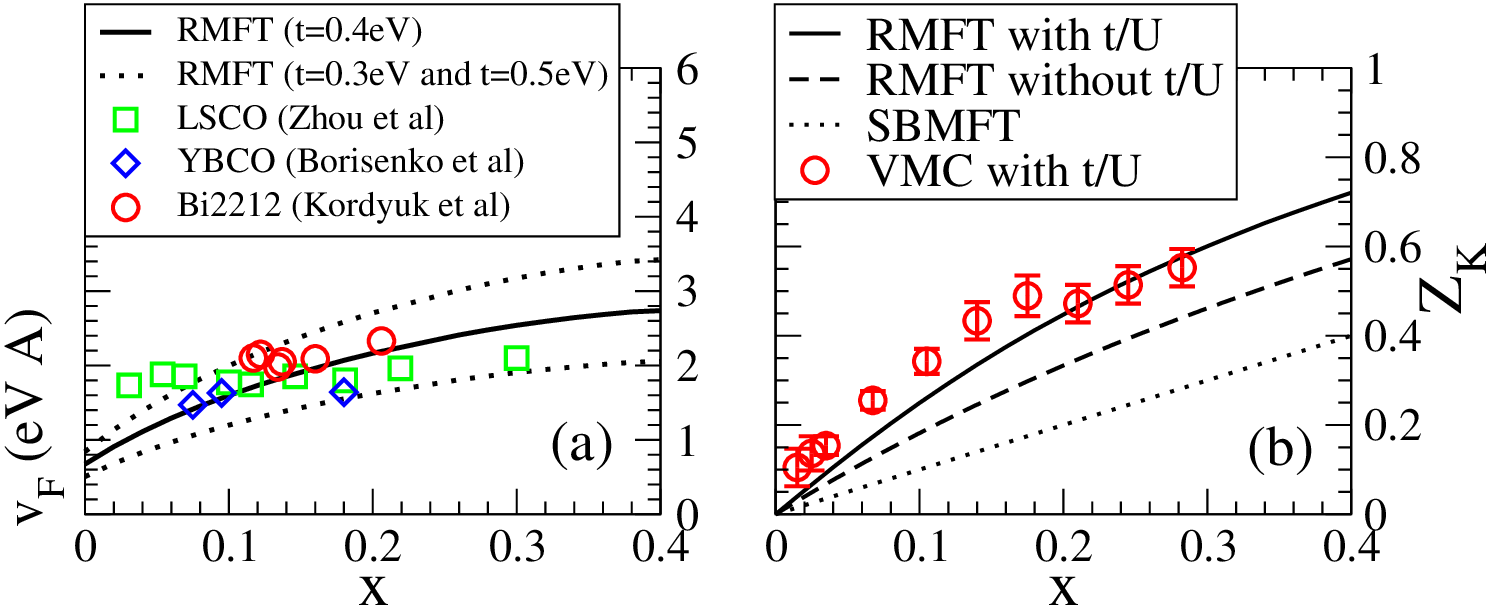}
 \caption{(color online) Doping dependence of (a) Fermi velocity, $v_F$,
and (b) renormalization, $Z_k$, of
the Gutzwiller-Bogoliubov nodal quasiparticle.
RMFT results are compared with experiments
(in (a), data from Ref.~\onlinecite{vFARPES}) and VMC
(in (b), data from Ref.~\onlinecite{paramekanti}), respectively.
%
%
\label{nodalpoint}
}
\end{figure}

We now consider the nature of the low lying excitations, the
quasiparticles created at the nodal point, $k_F$. The nodal dispersion
around $k_F$ is characterized by the velocity,
$v_F$, which directly influences a number of experimentally
accessible quantities. Within RMFT,
$v_F$ is directly obtained by calculating the gradient of
$\zeta_k$ along the direction, $(0,0) \rightarrow (\pi,\pi)$. The
result is presented in \fig{nodalpoint}(a) (for $t=0.3,0.4,0.5\ \text{eV}$ and $a_0=4 \AA$)
and is well approximated by the formula,
\begin{equation}
v_F/a_0 \approx \sqrt 2 \sin k_F \left[2g_t(t+2t^\prime \cos k_F)
+ x_1\frac{J}{4}\xi \right]~. \label{vf}
\end{equation}
In the above equation, we set $J'$, $J''$, and $x_D$ to zero
for simplicity. As seen in \fig{nodalpoint}, $v_F$ increases
with $x$, but remains finite as $x \rightarrow 0$. As can be
inferred from \eq{vf}, the energy scale of the
nodal velocity at $x=0$ is determined by $J$, \emph{i.e.},
${v_F}/({a_0
J})\approx\sqrt{2}\sin(k_F) \frac {11} 4 \xi \approx 1.5$ (for
$\xi\approx0.38$ and $k_F\approx \frac \pi 2$). The observed
doping dependence stems from the effects of Gutzwiller projection
$P_G$. As $x$ increases, holes gain kinetic energy by direct
hopping, \textit{viz.}, $g_t$ increases with doping; but $g_s$
decreases, leading to the doping dependence of $v_F$ seen in
\fig{nodalpoint}.

Our results agree with the numerical VMC results of
Paramekanti \textit{et al.}, who extract $v_F$ from the
discontinuity of the first moment of the spectral function in the
repulsive $U$ Hubbard model \cite{paramekanti}, and those of
Yunoki \textit{et al.}, who obtain $v_F$ from the quasiparticle
dispersion in the $t-J$ model \cite{yunoki_05}. A comparison to
ARPES data \cite{vFARPES}, presented in \fig{nodalpoint}(a), shows good agreement.
The doping dependence of $v_F$ in the severely underdoped regime remains
to be settled experimentally. While some groups
report a nearly constant Fermi velocity (see data for LSCO in \fig{nodalpoint}),
others observe an increase with doping (see data for YBCO and Bi2212 in \fig{nodalpoint}).
Within RMFT, we also verify that the nodal
properties remain essentially unchanged when $\Delta$ is
set to $0$; \textit{i.e.,} the doping dependence of $v_F$ results
from the vicinity of the state to a Mott insulator, rather than
the occurrence
of superconductivity itself.

\begin{figure}
\centering
\includegraphics[width=0.4\textwidth]{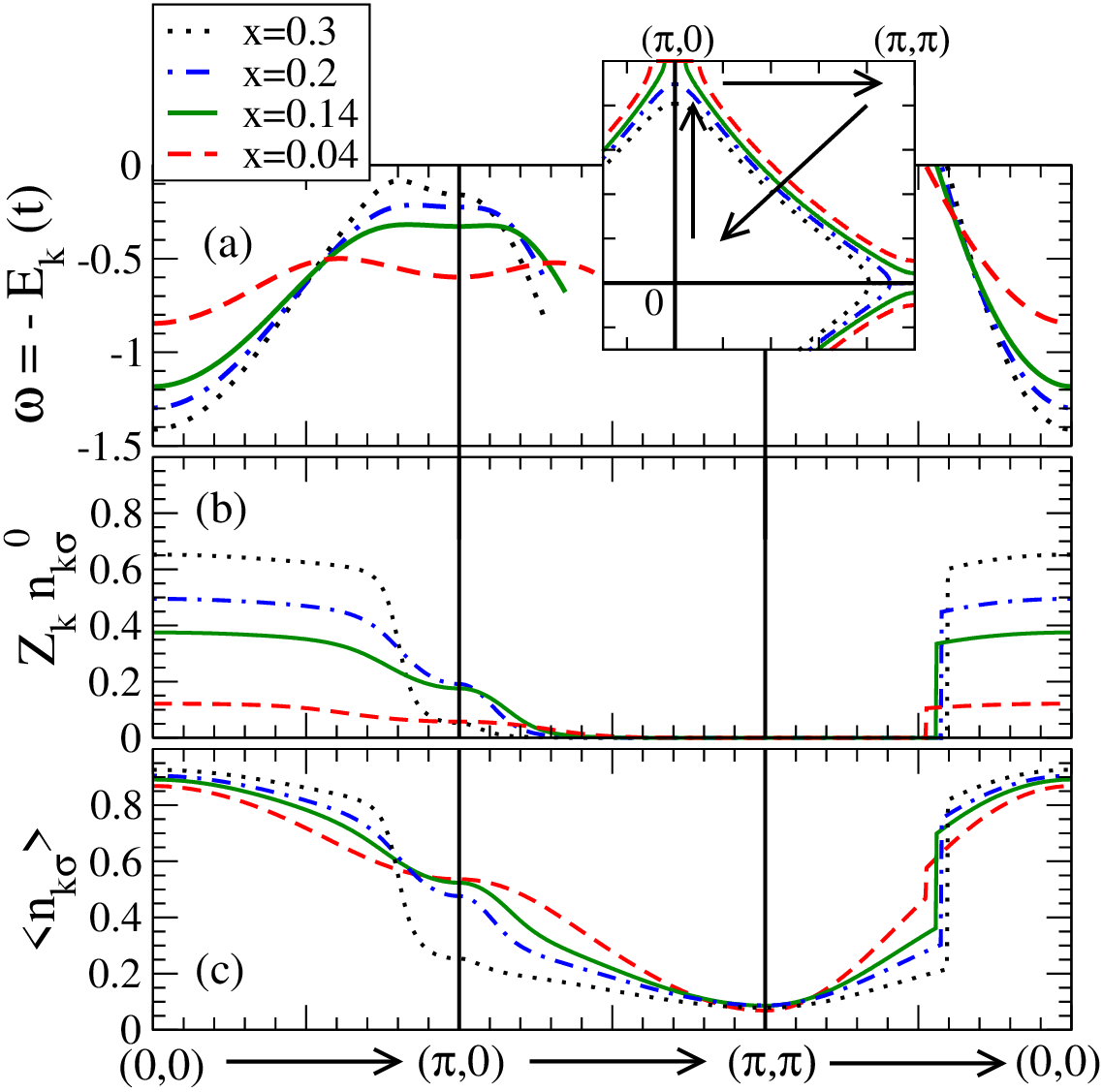}
\caption{(color online) (a) Energy dispersion, $\omega=-E_k$, (b)
quasiparticle weight, $Z_k n_{k\sigma}^0$, and (c) momentum
distribution, $\langle n_{k\sigma} \rangle$, of the
Gutzwiller-Bogoliubov quasiparticle for different doping $x$; The
energy dispersion [in (a)] is only shown when the corresponding
quasiparticle weight is finite [see (b)]. The corresponding Fermi
surface, $\zeta_k=0$, is shown in the inset of (a).
\label{qpenergy} }
\end{figure}
We now calculate the spectral weight of the Gutzwiller-Bogoliubov
quasiparticle (QP). The variational excited state in a projected
superconductor is given by $\exp(iS)|\Psi_{k}^-\rangle
\,\equiv\,\exp(iS) P \gamma^\dagger_{-k\downarrow}|\Psi_0\rangle$,
where the corresponding Bogoliubov QP operator is defined by
$\gamma^\dagger_{-k\downarrow}\equiv
u_kc^\dagger_{-k\downarrow}+v_k c_{k\uparrow}$. In ARPES, the
spectral weight corresponding to this excitation is determined by
the matrix element $M_{k}^- \equiv | \langle \Psi_{k}^-| \tilde
c_{k\uparrow}^{\phantom{\dagger}} |\Psi\rangle |^2 /( N_{k}^-
N_G)$, where, $N_{k}^-$ and $N_G$ are the norms of $|\Psi^-_k
\rangle$ and $|\Psi \rangle$, respectively. Here, $\tilde
c_{k,\sigma} =\exp(-iS) c_{k,\sigma} \exp(iS)$. Using the the
Gutzwiller renormalization scheme we find, $M_{k}^- \approx
Z_k\,n^0_{k\sigma}+\mathcal{O}(t/U)^2$ \cite{fn1} and,
\begin{equation}
Z_k\approx g_t+ \frac{g_3} U \left(\frac{1-x^2}2 \epsilon^0_{k} +
\frac{3-x} L \sum_{k'} v^2_{k'} \epsilon^0_{k'}\right) \ ,
\label{RMweight}
\end{equation}
for the QP renormalization \cite{fn2}, with $\epsilon^0_k=2 t
(\cos k_x+\cos k_y) + 4\,t' \cos k_x \cos k_y$. Here,
$n^0_{k\sigma}=v^2_k$, is the momentum distribution in the
unprojected wave function $|\Psi_0\rangle$. The renormalization,
$Z_k$, of the nodal QP weight is plotted as a solid line in
\fig{nodalpoint}(b), and agrees well with VMC results
\cite{paramekanti}. The dashed lines correspond to results without
$t/U$-corrections. The dotted line, $Z_k = x$, is the result from
slave boson mean field theory (SBMFT).

As a qualitative comparison with ARPES we show the
energy dispersion, $\omega=-E_k$, of the Gutzwiller-Bogoliubov QP
along the directions, $(0,0)\rightarrow(\pi,0)$,
$(\pi,0)\rightarrow(\pi,\pi)$, and $(\pi,\pi)\rightarrow(0,0)$ for
different $x$ in \fig{qpenergy}. We emphasize that our
calculations describe the low energy sector, and do not seek to
explain the ``kink'' at higher energies \cite{vFARPES}.

The spectral weight of the coherent peak, measured in ARPES, is
related to the QP weight, $M^-_k\approx Z_k n^0_{k \sigma}$; it is
shown in \fig{qpenergy}(b). As seen in the figure, the QP spectral
weight is severely modified by Gutzwiller projection. It decreases
with doping, and vanishes at half filling. This causes a shift of
spectral weight to an incoherent background as seen in the
momentum distribution function, $\langle n_{k\sigma}\rangle\approx
Z_k\,v^2_k+n^{\rm inc}_{k\sigma}+\mathcal{O}(t/U)^2$. While the
first term corresponds to the coherent QP weight, the second gives
the distribution of the incoherent part. We get,
\begin{align}
&n^{\rm inc}_{k\sigma}\approx\frac{(1-x)^2}{2(1+x)}\,+\sum_{\tau}
\frac {t_{\tau}} {2U} \cos(k \tau) \left[\frac{(1-x)^3}{1+x} +
\left(\frac {3g_s+1} {2} \right. \right. \nonumber\\
 &\quad- \left.g_3 \frac{3+x}2 \right) |\Delta_\tau|^2
\left.+\left(\frac {3g_s-1}{2}-g_3 \frac{3-x}2 \right) \xi^2_\tau
\right]~,
\end{align}
which is a smooth function of $k$. Results are shown in
\fig{qpenergy}(c) .
The incoherent weight is spread
over the entire Brillouin zone, and overlies the coherent part
from the Gutzwiller-Bogoliubov quasiparticles. At
half-filling, all weight becomes incoherent.

\begin{figure}
  \centering
 \includegraphics*[width=0.4\textwidth]{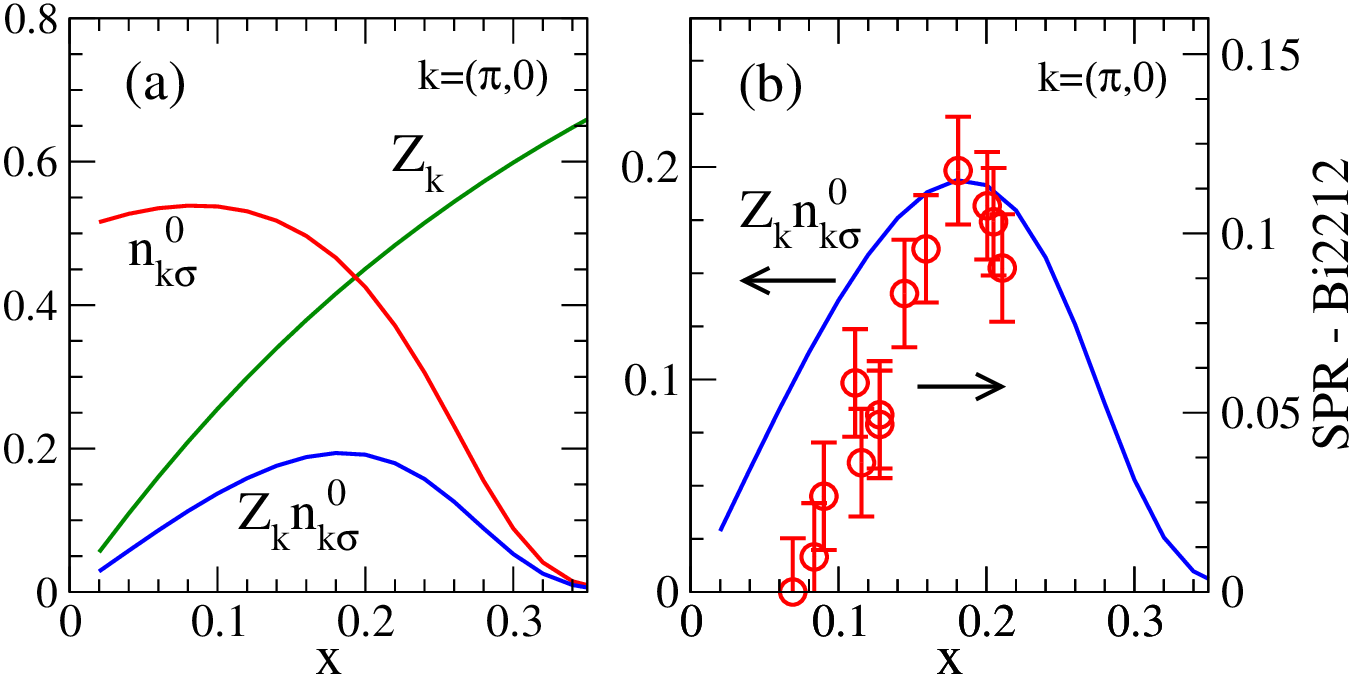}
 \caption{(color online) Doping dependence at the antinodal point, ${\mathbf k}=(\pi,0)$:
(a) QP renormalization, $Z_k$, the unrenormalized QP weight,
$n^0_{k\sigma}=v^2_k$, and
 the renormalized coherent QP weight, $Z_k v^2_k$; (b) coherent weight,
$Z_k\,v^2_k$ compared with the experimentally determined
Superconducting Peak Ratio (SPR) for Bi2212 \cite{feng_00}.}\label{pi0}
\end{figure}

Finally, we consider the coherent QP weight, $M^-_k\approx Z_k
v^2_k$, at the antinodal point, ${\mathbf k}=(\pi,0)$. As seen in
\fig{qpenergy}(b) and \fig{pi0}, it exhibits a non monotonic
behavior as a function of doping. Within our theory, this effect
arises from a combination of the effects due to Gutzwiller
projection and the topology change [see insert of
\fig{qpenergy}(a)]  of the underlying Fermi surface (FS).
\fig{pi0}(a) illustrates this clearly. While the QP weight
renormalization, $Z_k$, increases with increasing doping,
$n^0_k=v^2_k$, decreases due to the topology change, which occurs
at $x\approx0.15-0.20$ for our choice of hopping parameters
($t'=-t/4$). The change of the FS seems to be a generic feature of
hole doped cuprates \cite{topology}, although the exact doping
concentration $x$, for which this occurs, is sensitive to the
ratio between various hopping parameters. The combined effect of
strong correlations and topology change leads to a maximum of the
QP weight for the doping level, $x$, at which the underlying FS
changes topology. This result should be tested by appropriate
analyses of recent ARPES data \cite{topology}. Indications for
such a behavior have already been published
\cite{feng_00,ding_01}. Feng \textit{et al.} \cite{feng_00}
extracted the Superconducting Peak Ratio [SPR, illustrated in
\fig{pi0}(b)] which is proportional to the coherent QP spectral
weight, $Z_k\,v^2_k$. They found that the SPR increases with small
$x$, attains a maximum value around $x \approx 0.2$ where it
begins to decrease. Ding \textit{et al.} \cite{ding_01}, reported
similar results from ARPES. Although the topology change does not
influence the stability of the SC state within RMFT, the SC
pairing parameter $\Phi$ (related to $T_c$) and the QP weight,
$Z_k v^2_k$,  show some similarity as a function of doping.

To summarize, we studied the electronic structure of a Gutzwiller
projected $d$-wave superconductor. A systematic combination of
renormalized mean field theory and direct evaluation of matrix
elements was applied to the Hubbard model in the strong coupling
limit. Our analytical results can be used to fit experimentally
observed quantities as well as those obtained from numerical
methods. The dispersion of the Gutzwiller projected superconductor
is renormalized in the vicinity of the Mott insulator, but the
nodal Fermi velocity stays finite as the insulating limit is
approached. The spectral weight of the nodal quasiparticle
increases with doping, whereas that of the antinodal excitation
shows non monotonic behavior, when the underlying Fermi surface
changes topology. Our results can be checked by experimental
observations from ARPES of high temperature superconductors, thus
providing a way to study the applicability of projected wave
functions to these systems. The method we use is also amenable to
other extensions such as coupling to the lattice,
antiferromagnetism and long range interactions.

We thank T.~Valla, P.~D.~Johnson, J. Fink, and A.~A.~Kordyuk
for discussions. VNM
acknowledges partial support from the PSC-CUNY Research Award
Program. BE thanks the Hermann Willkomm Stiftung for supporting his
stay in Princeton.


\begin{thebibliography}{99}
\bibitem{calc_energy} Since the calculation of the expectation
value, $\langle \tilde{H}_{{\rm eff}}\rangle_0$, can be found in
Ref.~\onlinecite{edegger_05}, and the derivation of the gap equations
follows Ref.~\onlinecite{zhang_88}, we omit details
here.
\bibitem{fn1} Here, we dropped a small contribution to
 $M^-_k$ of $\mathcal{O}(t/U)$, which cannot be written as $Z_k n^0_{k\sigma}$.
\bibitem{fn2} A similar result with minor discrepancies, was obtained independently in
Ref.~\onlinecite{randeria_05}.
\end{thebibliography}
\end{document}